\documentclass[twocolumn]{svjour3}          

\usepackage{graphicx}                       
\usepackage{amsmath,amsfonts,amssymb}       
\usepackage{hyperref}                       
\usepackage{float}                          
\usepackage{lipsum}                         
\usepackage{newtxtext}                      
\usepackage{newtxmath}                      
\usepackage{multirow}                       
\usepackage[title]{appendix}                
\usepackage{xcolor}                         
\usepackage{textcomp}                       
\usepackage{booktabs}                       
\usepackage{algorithm}                      
\usepackage{algpseudocode}                  
\usepackage{listings}                       
\usepackage[nolist]{acronym}
\journalname{Springer Journal}              
\usepackage{comment}                        
\usepackage[most]{tcolorbox} 

\begin{acronym}
    \acro{slp}[SLP]{Software Product Lines}
    \acro{cse}[CSE]{Continuous System Engineering}
    \acro{wp}[WP]{work packages}
    \acro{pr}[PR]{Pull request}
    \acro{llm}[LLM]{Large Language Models}
    \acro{kpi}[KPI]{Key Performance Indicators}
    \acro{rag}[RAG]{Retrieval-Augmented Generation}
    \acro{mvp}[MVP]{Minimally Viable Product}
\end{acronym}

\tcbset{
    mybox/.style={
        colframe=black, 
        colback=white, 
        boxrule=0.4pt, 
        arc=5pt, 
        width=\linewidth, 
        left=0pt, 
        right=2pt, 
        top=2pt, 
        bottom=2pt 
    }
}

\title{SmartDelta Methodology: Automated Quality Assurance and Optimization for Incremental System Engineering}

\author{Benedikt Dornauer\textsuperscript{1*} \and 
        Michael Felderer\textsuperscript{1,2} \and 
        Mehrdad Saadatmand\textsuperscript{3}  \and
        Muhammad Abbas\textsuperscript{3}  \and
        Nicolas Bonnotte\textsuperscript{4}  \and 
        Andreas Dreschinski\textsuperscript{4} \and 
        Eduard Paul Enoiu\textsuperscript{5} \and 
        Eray Tüzün\textsuperscript{6} \and 
        Baykal Mehmet Uçar\textsuperscript{7} \and 
        Ömercan Devran\textsuperscript{7} \and 
        Robin Gröpler\textsuperscript{8}
        }

\institute{
    \textsuperscript{1} University of Innsbruck, Austria\\
    \textsuperscript{2} German Aerospace Center (DLR), Institute of Software Technology, Cologne, Germany\\ 
    \textsuperscript{3} RISE Research Institutes of Sweden, Västerås, Sweden\\
    \textsuperscript{4} Akkodis Germany Solutions GmbH, Sindelfingen, Germany\\ 
    \textsuperscript{5} Mälardalen University, Västerås, Sweden\\ 
    \textsuperscript{6} Bilkent University, Ankara, Turkey\\ 
    \textsuperscript{7} Arçelik, Istanbul, Turkey\\ 
    \textsuperscript{8} ifak Institute for Automation and Communication, Magdeburg, Germany\\ 
    *Corresponding authors. E-mails: 
    \href{mailto:benedikt.dornauer@uibk.ac.at}{benedikt.dornauer@uibk.ac.at}; 
    \\
    Contributing authors: \href{mailto:mehrdad.saadatmand@ri.se}{mehrdad.saadatmand@ri.se} 
    \href{mailto:michael.felderer@dlr.de}{michael.felderer@dlr.de}; \href{mailto:muhammad.abbas@ri.se}{muhammad.abbas@ri.se};  
    \href{mailto:nicolas.bonnotte@akkodis.com}{nicolas.bonnotte@akkodis.com}; 
    \href{mailto: dreschinski@akkodis.com}{andreas.dreschinski@akkodis.com}; 
    \href{mailto: paul.enoiu@mdu.se}{eduard.paul.enoiu@mdu.se}; 
    \href{mailto:baykal.ucar@beko.com}{baykal.ucar@beko.com};
    \href{mailto:omercan.devran@beko.com}{omercan.devran@beko.com};
    \href{mailto:eraytuzun@cs.bilkent.edu.tr}{eraytuzun@cs.bilkent.edu.tr}; \href{mailto:robin.groepler@ifak.eu}{robin.groepler@ifak.eu}
    }
\date{}

\begin{document}

\maketitle

\begin{abstract}
     Modern software systems undergo frequent updates, continuously evolving with new versions and variants to offer new features, improve functionality, and expand usability. Given the rapid pace of software evolution, organizations require effective tools and methods to mitigate the challenges associated with these changes, also called deltas. To address these challenges, the international SmartDelta Project joined industry and academia to develop and test solutions for incremental development and quality assurance. This paper provides insights into the SmartDelta project achievements and highlights one main contribution: the \textit{SmartDelta Methodology}, a domain-unspecific concept for delta management in incremental software engineering. This methodology enables companies to identify gaps in their continuous engineering environment across six stages and helps to discover new tools in various technical areas. Additionally, the paper presents seven selected tools at different stages of the methodology.
\end{abstract}

\section{Introduction}
\label{intro}
    Software undergoes continuous evolution. These changes occur through iterative development, where, for instance, new features are incrementally added, anomalies are identified and fixed, or performance optimization is conducted. This often leads to new software versions or variants. While updates may enhance certain quality aspects, other aspects can deteriorate over time \cite{Li2015-uu}. For instance, trade-offs can arise between security and performance or between feature expansion and code quality, especially in the long term.
    
    \begin{table}[ht]	
\vspace{-0.3cm}
\centering
\caption{SmartDelta Industrial Use Cases}
\label{tab:useCases}

\begin{tabular}{p{0.6cm} p{7cm}}
\toprule
\textbf{Ctry.} & \textbf{Company and Industry Sector} \\
\hline
\raisebox{-0.6\height}{\frame{\includegraphics[width=0.45cm]{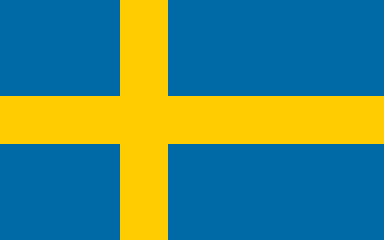}}} & Alstom (Railway) \\ 
\cline{2-2}
\raisebox{-0.6\height}{\frame{\includegraphics[width=0.45cm]{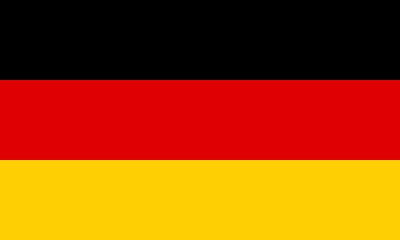}}} & Akkodis (eMobility), Software AG (Enterprise Software)\\
\cline{2-2}
\raisebox{-0.6\height}{\frame{\includegraphics[width=0.45cm]{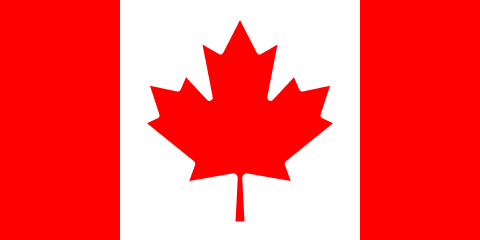}}} & eCAMION (eMobility), GlassHouse Systems (Cybersecurity) \\ 
\cline{2-2}
\raisebox{-0.6\height}{\multirow{2}{*}{\frame{\includegraphics[width=0.45cm]{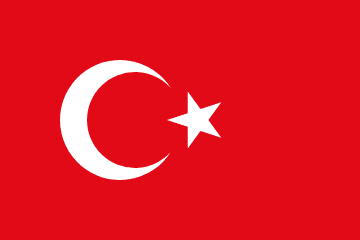}}}} & NetRD (Telecommunication), Kuveyt Türk Bank (Banking and Finance), Arçelik (Home Appliances) \\ 
\cline{2-2}
\raisebox{-0.6\height}{\frame{\includegraphics[width=0.45cm]{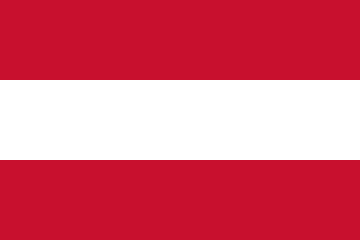}}} & c.c.com Moser GmbH (Logistics and Personal Mobility) \\ 
\cline{2-2}
\raisebox{-0.6\height}{\frame{\includegraphics[width=0.45cm]{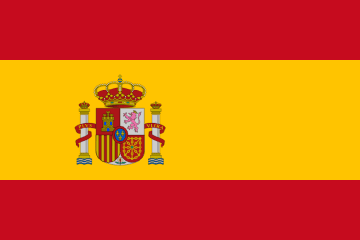}}} & Izertis (Enterprise Software) \\ 
\cline{2-2}
\raisebox{-0.6\height}{\frame{\includegraphics[width=0.45cm]{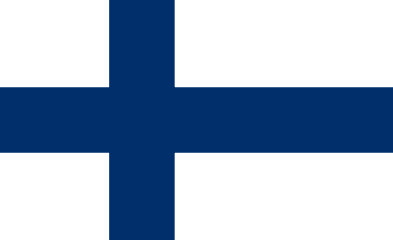}}} & Vaadin (Software Development Platform) \\ 
\bottomrule
\end{tabular}
\vspace{-0.3cm}
\end{table}

        \begin{figure*}[ht]
        \centering
        \includegraphics[width=0.9
        \linewidth]{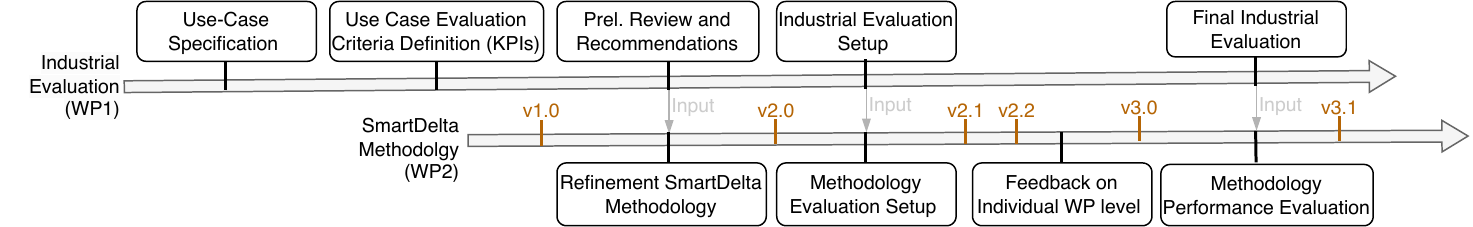}
        \caption{Step-by-step industrial evaluation setup and contribution to the Methodology}
        \label{fig:WP1Steps}
        \vspace{-0.4cm}
    \end{figure*}
    \begin{figure}[ht]
        \centering
        \includegraphics[width=0.9\linewidth]{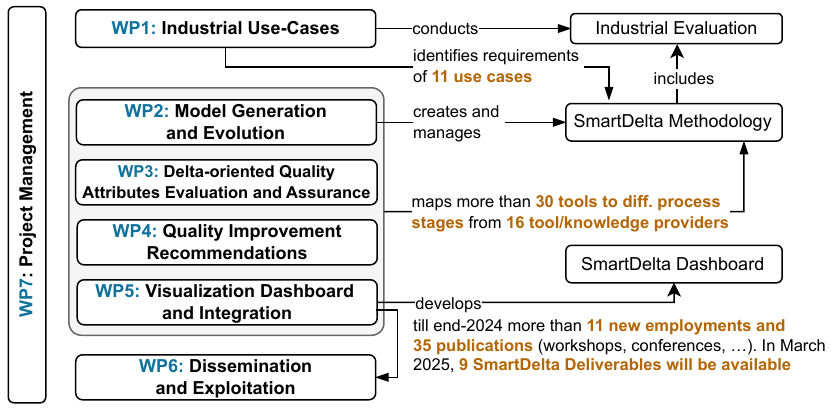}
        \caption{SmartDelta Project Working Structure}
        \label{fig:workPackages}
        \vspace{-0.5cm}
    \end{figure}
    
    To consolidate problem statements, business objectives, and challenges related to incremental systems engineering, 11 use cases across various industry domains (see Table \ref{tab:useCases}) contributed their perspectives and requirements. The challenges identified, provided in detail in \cite{SmartDeltaJournal}, underscore the importance  of precise analysis and assessments of the adjustments associated with each modification to a system, which are crucial in continuous engineering. Thus, the goal of SmartDelta\footnote{\url{https://itea4.org/project/smartdelta.html}} has been to build automated solutions for quality assessment of product deltas in a continuous engineering environment. It provides intelligent analytics from development artifacts (e.g., source code, log files, requirement specifications) and system execution, offering insights into quality improvements or degradation resulting from various changes, along with recommendations for the next build.

    To address the various delta-management aspects, the project contains different \ac{wp}, each focusing on a specific topic. Specifically, \ac{wp} 2 focuses on models for generation and evolution, while \ac{wp} 3 concentrates on tools for delta-oriented quality assurance, and \ac{wp} 4 develops solutions for quality improvement recommendations. Since most tools integrate visual representations of various delta aspects, \ac{wp} 5 supports the development of guidelines and approaches for the visualization part. 
    
    Managing the work packages' different outcomes (knowledge, solutions, tools, etc.) and organizing these contributions within a coherent structure is challenging, especially for such a large consortium. Consequently, we developed the \textit{SmartDelta Methodology}, a novel continuous engineering management concept that emerged through an incremental design process illustrated in \ref{fig:WP1Steps}. This process involved a validity evaluation conducted by the use case providers, who offered recommendations to enhance the SmartDelta Methodology further. Apart from that, the use cases tried and tested the various solutions in \ac{wp} 1.
    
    The structure of the paper is as follows: Section \ref{sec:methodology} discusses the term of delta in software engineering and presents the SmartDelta Methodology. Section \ref{sec:tools} discusses solutions associated with specific process stages. Finally, Section \ref{sec:conclusion} provides a short recap and outlines the future work.

\section{Delta Management Concept: The \texorpdfstring{\NoCaseChange{SmartDelta}}{SmartDelta} Methodology}
\label{sec:methodology}
    \begin{figure*}[ht]
        \centering
        \includegraphics[width=0.9\textwidth]{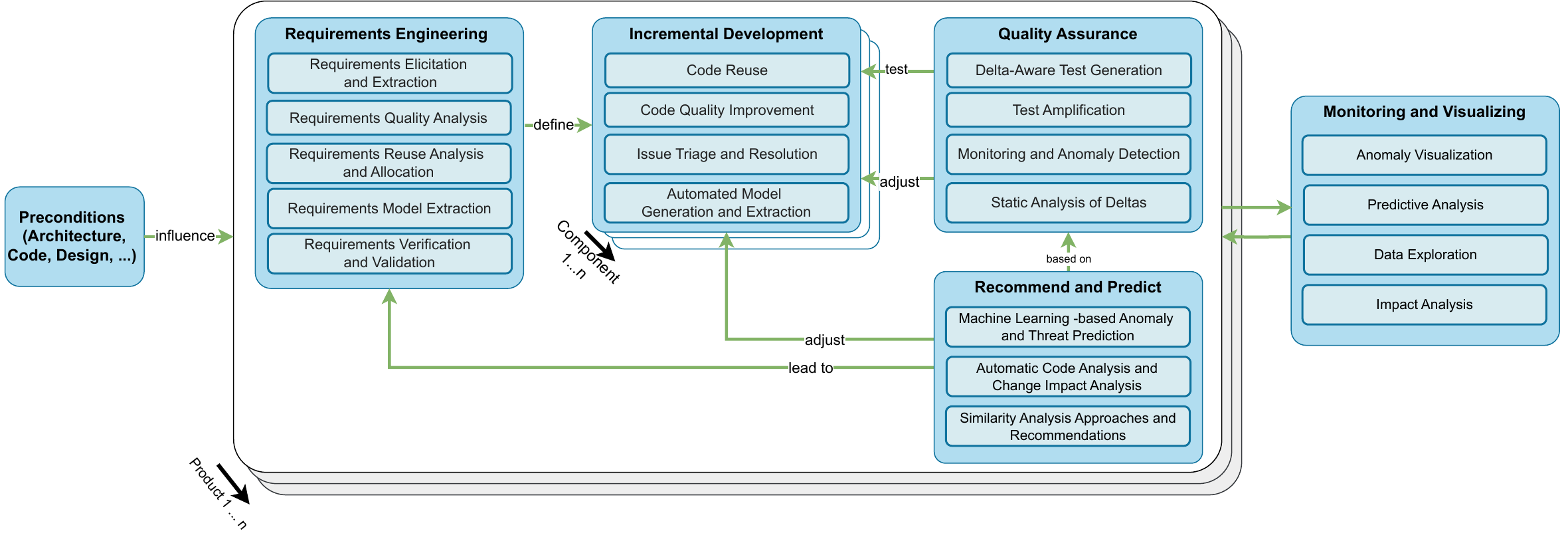}
        \caption{The SmartDelta Methodology}
        \label{fig:smartDeltaMethodology}
        \vspace{-0.4cm}
    \end{figure*}

    The concept of delta has been explored in areas like software configuration management, version control, product line engineering, and variability management (e.g., \cite{haber2012evolving,lochau2012incremental,seidl2019software,linsbauer2018variability}). Deltas are defined differently across these domains, tracking changes between versions in configuration management while representing variations in product lines. 
    In SmartDelta, a \textit{Delta} is any change in a software product that results in a new product instance with different quality and/or functionality properties. Examples of such changes include adding or removing features and components, applying fixes and updates, reconfiguring a software product for deployment in a new environment, or adapting to the needs of a new customer. 
    
    This concept is often linked to the temporal aspects of product evolution and is commonly referred to as a \textit{Version Delta}. A version represents a specific release of a software product. \textit{A Variant Delta} refers to a distinct form of software modified or customized for specific purposes or applications. These variants may differ in functionality, features, or configurations while maintaining a common core derived from the original software \cite{10.1145/280277.280280}. 

    To manage these kinds of deltas, we created the management concept called the \textbf{\textit{SmartDelta Methodology}}, illustrated in Figure \ref{fig:smartDeltaMethodology}. A primary goal of the SmartDelta Methodology is to offer a holistic overview guiding companies in their (individual) incremental software systems development. Therefore, the concept consists of six stages: 

    \smallskip
    
    \noindent The iterative development process is driven by certain \textbf{Preconditions}: pre-existing artifacts, such as architecture, code, and design, which serve as both constraints and guidance. These artifacts, alongside input from stakeholders and other external influences, determine the development direction of a product.

    \smallskip
    
   \noindent \textbf{Requirements Engineering} is structured as an iterative, delta-focused process aimed at defining and adapting the necessary project requirements over time. Requirements are continuously identified and adapted from diverse sources, with a strong focus on tracking and analyzing each delta (i.e., any modification or adjustment needed to meet evolving goals). Quality standards are ensured through delta-aware quality analysis, where existing requirements are analyzed for reuse, adapting them efficiently to new contexts and minimizing redundant efforts. Model extraction plays a crucial role in visualizing requirements and their deltas to understand the impact of each change on the system as a whole. Verification and validation processes confirm that these evolving requirements align with stakeholder needs and adhere to system constraints.

    \smallskip
    
    \noindent In the \textbf{Incremental Development} phase, software systems are constructed using reusable, modular components that address the requirements. By doing so, integrating feedback from quality assurance is essential to identify and resolve potential issues early in the development cycle. This iterative process ensures that each increment aligns with the overall system objectives while maintaining high-quality standards. From a product line management perspective, the increment process begins with the development of a usable, improvable product that may not be an \ac{mvp} initially but evolves through feedback until it reaches \ac{mvp}. This iterative build-measure-learn cycle, which eliminates undesirable deviations, is applied across different versions and variants, ensuring reusability and continuous improvement in product development.

   \smallskip
    
   \noindent Delta-Driven \textbf{Quality Assurance} highlights testing, validation, and review. Each change, whether a minor code adjustment or a significant feature update, is rigorously evaluated to ensure quality standards are met throughout development. Delta-aware test generation targets specific changes, ensuring all modifications meet defined requirements and quality standards. Test amplification improves this by expanding test cases to cover scenarios introduced by deltas. Continuous monitoring and anomaly detection provide real-time feedback, highlighting any quality degradation linked to changes. Additionally, static analysis of deltas assesses modifications against predefined quality metrics. Integrating these delta-aware techniques enables effective defect detection at this stage. Quality assurance and test engineers apply these techniques to manage and execute the delta-aware quality assurance process directly. Software developers play an essential role in implementing feedback to support continuous improvement, while project managers ensure that quality goals set using delta-driven quality assurance align with project timelines and standards.

    \smallskip
    
    \noindent The \textbf{Recommend and Predict} stage leverages delta-aware analysis to generate predictive recommendations for software evolution, often utilizing AI. By analyzing data from the quality assurance phase, this stage identifies areas needing adjustments, frequently resulting in new requirements and highlighting reusable components for future developments. This delta-focused approach ensures that each prediction and recommendation is backed by data-driven insights, allowing the system to adapt as it evolves.  
    
    \smallskip
    
    \noindent Last but not least we have \textbf{Monitoring and Visualizing}. It involves tracking and visualizing key metrics, states, dependencies, and interactions across all system components and the system itself. Therefore, data from previous stages is used to monitor development progress, requirements status, and quality through real-time dashboards, identifying trends, bottlenecks, and alignment with objectives, thus supporting data-driven decisions and continuous improvement.
    
\section{A Glimpse of \texorpdfstring{\NoCaseChange{SmartDelta}}{SmartDelta} Tools}
    \label{sec:tools}
    For the stages, multiple tools have emerged. They can be mapped to different technical areas for each stage, provided in \ref{fig:smartDeltaMethodology}. From a corpus of 33 tools, we have selected seven as demonstrators.

    \begin{table}[ht]
\centering
\caption{SmartDelta Tool and Knowledge Providers}
\label{tab:toolProvider}

\begin{tabular}{p{0.3cm} p{7cm}}
\toprule
\textbf{Ctry.} & \textbf{Tool and Knowledge Provider} \\
\hline
\raisebox{-0.6\height}{\frame{\includegraphics[width=0.45cm]{figures/flags/se.png}}} & Mälardalen University, RISE - Research institutes of Sweden\\ 
\cline{2-2} 
\multirow{2}{*}{\raisebox{-.4\height}{\frame{\includegraphics[width=0.45cm]{figures/flags/de.png}}}} & Cape of Good Code,  Fraunhofer FOKUS, IFAK Institute for Automation and Communication, TWT GmbH Science \& Innovation, Akkodis\\ 
\cline{2-2} 
\frame{\raisebox{-.5\height}{\includegraphics[width=0.45cm]{figures/flags/ca.png}}} & Cyberworks Robotics, Team Eagle, Ontario Tech University \\ 
\cline{2-2} 
\raisebox{-.4\height}{\frame{\includegraphics[width=0.45cm]{figures/flags/tr.png}}} & ERSTE Software Limited, Bilkent University, Arçelik\\ 
\cline{2-2} 
\raisebox{-.4\height}{\frame{\includegraphics[width=0.45cm]{figures/flags/at.png}}} & University of Innsbruck\\ 
\cline{2-2} 
\raisebox{-.4\height}{\frame{\includegraphics[width=0.45cm]{figures/flags/es.png}}} & University of Madrid Carlos III\\ 
\cline{2-2} 
\raisebox{-.4\height}{\frame{\includegraphics[width=0.45cm]{figures/flags/fi.png}}} & Hoxhunt\\
\cline{2-2} 
\raisebox{-.4\height}{\frame{\includegraphics[width=0.452cm]{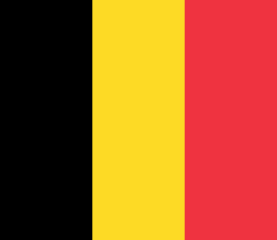}}} & University of Antwerp \\
\bottomrule
\end{tabular}
\vspace{-0.3cm}
\end{table}

    \subsection{NALABS (NAtural LAnguage Bad Smells)}
        \begin{tcolorbox}[mybox]
            \begin{itemize}
                \item \textbf{Main Stage:} Requirements Engineering
                \item \textbf{Technical Area(s):} Requirements Quality Analysis, Requirements Verification 
            \end{itemize}
        \end{tcolorbox}
    
        NALABS is developed to automatically detect issues in natural language requirements and test specifications used in large-scale system development \cite{rajkovic2022nalabsdetectingbadsmells}. 
  
        Previous research \cite{juergens2010can} has shown that specifications written in natural language often contain significant cloning and structural issues, which can increase the cost of maintaining and executing test cases. For both requirements and system test cases written in natural language, specific "bad smells" have been identified as indicators of poorly structured artifacts. Building on these observations, we developed this set of indicators to detect flaws in requirements, defining dictionary-based metrics to automatically identify these bad smells in natural language artifacts.

        The tool identifies various bad smells or poor quality indicators in requirement documents, which can lead to costly errors and inefficiencies during development. NALABS uses dictionary-based metrics to detect issues such as vagueness, referenceability, optionality, subjectivity, weakness, readability, and complexity. 
        Available on GitHub\footnote{\url{https://github.com/eduardenoiu/NALABS}}, NALABS aims to improve the requirements engineering process by automating the detection of problematic requirements. 
        
\begin{figure}[ht]
            \includegraphics[width=\linewidth]{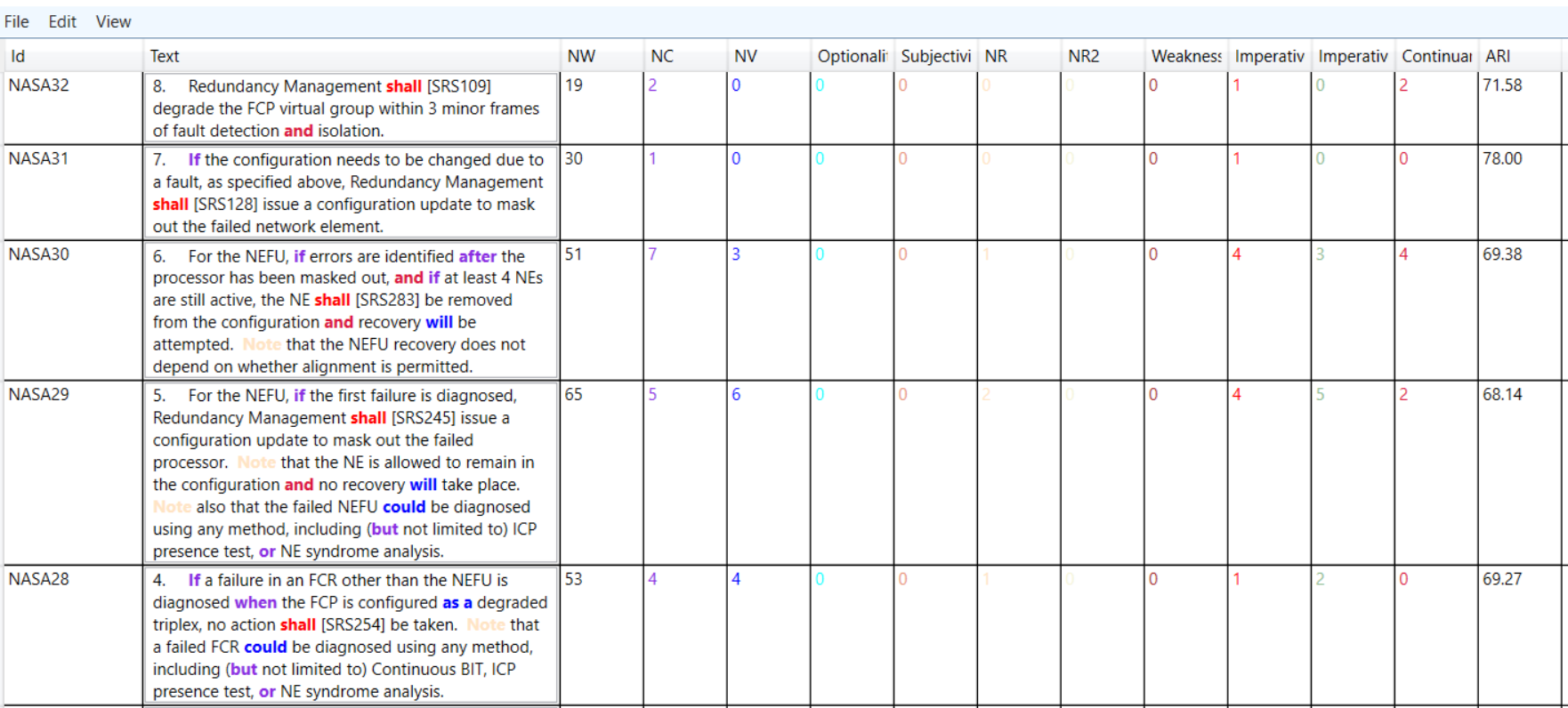}
            \caption{
              Usage of NALABS Tool from GUI.}
              \label{fig-context} 
              \vspace{-0.5cm}
        \end{figure}

        Figure \ref{fig-context} shows using NALABS from a GUI interface. The tool contains three essential layers: (i) the pre-processing of requirement documents stored as excel spreadsheets\footnote{Many companies and tools (such as IBM's Rational DOORS) use this format for working with requirements.}, (ii) the configuration and application of bad smell metrics, and (iii) presenting these results to the user. 
        NALABS is used in continuous development environments where requirements frequently evolve, as it enables automated detection of quality issues in new variants and deltas of requirements. NALABS helps prevent the propagation of errors and inefficiencies across different system versions in the Alstom use case. The tool detected ten issues for every 300 requirements and was evaluated on a dataset of 7,423 requirements from various projects. In addition, the NALABS tool analyzed 661 security requirements across 46 documents \cite{hallberg2023finding}, focusing on identifying "requirement smells"—indicators of potential issues such as vagueness, ambiguity, and excessive complexity that can hinder the usability of security requirements during incremental development. Certain indicators were statistically significant, pointing to common problem areas in security specifications. For example, ARI scores indicated a generally difficult to very difficult reading level across the requirements. Additionally, nine moderate to strong correlations were found among different smell indicators, revealing patterns in the language of security requirements. Notably, requirements with high conjunction counts often exhibited increased logical complexity, potentially obscuring intended functionality and raising re-implementation risks.

    \subsection{DRACONIS} 
        \begin{tcolorbox}[mybox]
            \begin{itemize}
                \item \textbf{Main Stage:}  Incremental Development
                \item \textbf{Technical Area(s):} Requirements Verification, Code Reuse, Code Quality Improvement 
            \end{itemize}
        \end{tcolorbox}

         Darconis is a framework designed to automate reviews in incremental block-based safety-critical software processes, particularly in the railway domain within the Alstom use case. It utilizes static round-trip analysis to enforce design rules and facilitate faster feedback at the model level. By translating models into an intermediate representation, DRACONIS enables metrics extraction and rule-based analysis of block-based programs like FBD and Simulink. The tool supports incremental analysis, checking only the modified parts of a model, thus improving review efficiency. DRACONIS also allows the creation of custom analysis rules using JSON files, making it adaptable to specific project requirements. This framework helps bridge the gap between developed models and their analyzed counterparts, ensuring adherence to safety standards and reducing manual review efforts in continuous development \cite{vassallo2020developers}.
        
        In addition to being a general analysis framework for block-based programs, DRACONIS is one of the tools in SmartDelta that automates the quality assurance process in an incremental development setting. Analyzing an intermediate representation allows the analysis of multiple models, and the analysis is presented in an accessible way. Detailed usage instructions, including small test models, can be found in a GitHub repository \footnote{https://github.com/jean-malm-mdh/draconis}. 
        
        In SmartDelta, DRACONIS was applied to use cases from the railway domain to evaluate its potential for accelerating design rule checks. An internal version of DRACONIS was tested on nine FBD models of varying complexity from an existing industrial project. When compared to a manual baseline, where an experienced developer performed the same analysis, DRACONIS demonstrated an average speedup of approximately 150 times across five executions.
     \subsection{SoHist v2}
            \begin{tcolorbox}[mybox]
                \begin{itemize}
                    \item \textbf{Main Stage: }  Quality Assurance
                    \item \textbf{Technical Area(s):} Code Quality Improvement, Static Analysis of Deltas, Automatic Code Analysis and Change Impact Analysis
                \end{itemize}
            \end{tcolorbox}
    
            SonarQube has some limitations regarding historical code analysis, including issues with version variability, a lack of filtering options, and inadequate visualization capabilities. To address these challenges, a tool called SoHist was developed and presented at EASE 2023 \cite{10.1145/3593434.3593460}. This tool enhances SonarQube's functionality by providing additional analysis features.
            \begin{figure}
                \centering
                \includegraphics[width=\linewidth]{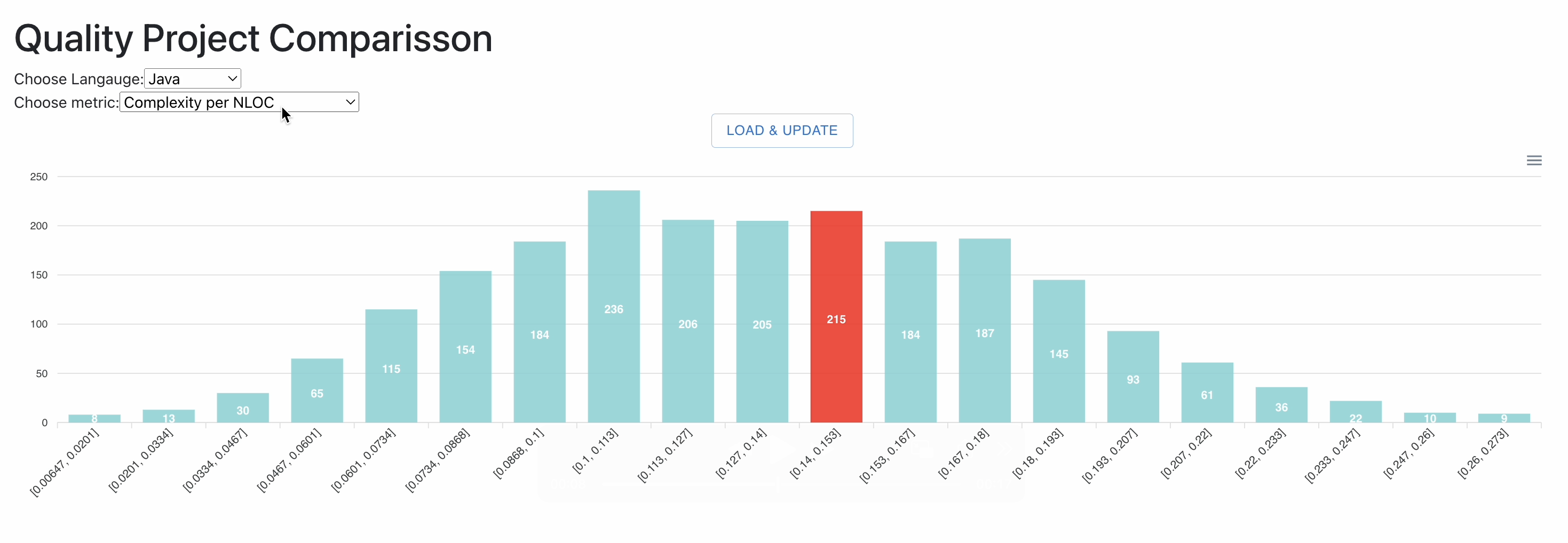}
                \caption{The red bar indicates the complexity density of the Java project compared to other projects.}
                \label{fig:sohist}
                \vspace{-0.3cm}
            \end{figure}
            In 2024, a new quality assessment feature was introduced. This feature utilizes an initial dataset comprising over 2007 historically evaluated SonarQube projects that meet specific quality criteria, such as having more than 100 commits, a minimum of 4 contributors, and at least 10 GitHub stars. With this data, users can benchmark their project's quality metrics against a large set of comparable projects, considering various programming languages and selected quality criteria. A chart visualizes the distribution of these metrics across other projects, yielding insights into metrics like Test Coverage, Code Smell Density, and more. An example of Code Complexity is given in Figure \ref{fig:sohist}. 
    \subsection{SmartMetrics}
        \begin{tcolorbox}[mybox]
            \begin{itemize}
                \item \textbf{Main Stage:} Incremental Development 
                \item \textbf{Technical Area(s):} Code Reuse
            \end{itemize}
        \end{tcolorbox}
        
         SmartMetrics analyzes files across repositories and evaluates software metrics like complexity and activity rates, and performs semantic analysis using topic modeling and embedding calculations (for \ac{rag}). It also supports smart artifact labeling by leveraging \ac{llm}. The tool supports delta analysis by comparing two versions (commits). Additionally, it creates interactive dashboards to visualize the analysis results, providing deep insights into code evolution and supporting \ac{rag} use cases. SmartMetrics has brought new insights into complex project environments, leading to improved decision-making. Furthermore, it enables joined evolution analysis of all artifacts, including requirements, which improves software quality in the long term.

    \subsection{Smellyzer} 
        \begin{tcolorbox}[mybox]
            \begin{itemize}
                \item \textbf{Main Stage: }  Quality Assurance 
                \item \textbf{Technical Area(s):} Issue Triage and Resolution, Monitoring and Anomaly Detection
            \end{itemize}
        \end{tcolorbox}
        
        Smellyzer is designed to elevate software development quality by identifying and addressing process smells within code review~\cite{codereviewprocesssmells} and bug tracking~\cite{bugprocesssmelssinPractice,bugtrackingprocesssmells}. By pinpointing detrimental patterns, such as large changesets, sleeping reviews, unassigned bugs, and the absence of traceability links to bug-fixing commits, Smellyzer provides teams with actionable insights to refine their practices. The latter issue, where bugs are closed without any linked corrective commit, has also been explored through semi-automated traceability tools like ReLink~\cite{relink}, which employ similarity-based heuristics to recover missing links between pull requests and reported issues.

        \subsection{PieR} 
            \begin{tcolorbox}[mybox]
                \begin{itemize}
                    \item \textbf{Main Stage: }Incremental Development  
                    \item \textbf{Technical Area(s):} Issue Triage and Resolution, Predictive Analysis  
                \end{itemize}
            \end{tcolorbox}
        
            PieR proposes an automated system to simplify the management of \ac{pr} by classifying them into specific categories with the help of \ac{llm}s. These categories, including "bug fix," "new feature," "security update," and others, address different aspects of software maintenance and development. This ensures \ac{pr}s are categorized and handled efficiently based on their content, type, and root causes. This tool enables project managers to learn about the distribution and nature of \ac{pr}s, encouraging informed decision-making and helping to understand project development trends better.	
        
        \subsection{Issue Analyzer} 
            \begin{tcolorbox}[mybox]
                \begin{itemize}
                    \item \textbf{Main Stage: }Recommend and Predict 
                    \item \textbf{Technical Area(s):} Similarity Analysis Approaches and Recommendations  
                \end{itemize}
            \end{tcolorbox}  
            Issue Analyzer performs automatic labeling and similarity analysis of software issues. This tool classifies issues, such as bug reports and feature requests, according to their label, e.g. their security relevance, with the aim of reducing the time required for manual issue identification. Using modern language models and vector databases, this tool also suggests similar issues to support bug fixing by utilising knowledge from previous bug fixes. This tool serves as a basis for further analyses, such as providing recommendations for code and test case reuse or assigning specific people or teams for design, implementation and testing. This tool improves the efficiency and accuracy of defect management and contributes to safe and maintainable software.

\vspace{-0.4cm}
\section{Conclusion and Future Work}

\label{sec:conclusion}
Software systems are typically not built and developed from scratch but rather as increments over existing versions. As such, over time, companies have a wide range of product versions and variants that need to be maintained and assessed for their quality and reusability. In this paper, we presented an update (January, 2025) of  the outcomes of SmartDelta, an international project developing solutions for managing software quality across different product versions and variants. In particular, the SmartDelta Methodology, along with a subset of project tools and solutions developed for implementing the methodology, were discussed. 

In total, we expect to have more than 30 solutions ready by the end of the project in March 2025. Due to space limitations, this paper provides a general overview of the project's outcomes. Detailed information is available in several public deliverables, which can be accessed in the \textit{Documents} section of the ITEA SmartDelta project webpage\footnote{\url{https://itea4.org/project/smartdelta.html}}. In addition to the deliverables, we will release a SmartDelta dashboard that visualizes examples of various delta aspects, plan to deploy a SmartDelta ChatBot enriched with knowledge from the Work Packages (utilizing a \ac{rag}) to enable quick access to relevant information, and publish an interactive web version of the SmartDelta Methodology.

\begin{acknowledgements}
This work was supported by and done within the scope of the ITEA3 SmartDelta project, which the national funding authorities of the participating countries have funded.
\end{acknowledgements}

\bibliographystyle{spmpsci} 
\bibliography{references.bib}

\end{document}